\documentclass[prl,twocolumn,showpacs]{revtex4}
\usepackage{graphicx}
\begin{document}

\title{Core Precession and Global Modes in Granular Bulk Flow}

\author{Denis Fenistein} \affiliation{Kamerlingh Onnes Lab,  Leiden
University, PObox 9504, 2300 RA Leiden, Netherlands}
\author{Jan-Willem van de Meent} \affiliation{Kamerlingh Onnes Lab,  Leiden
University, PObox 9504, 2300 RA Leiden, Netherlands}
\author{Martin van Hecke}\affiliation{Kamerlingh Onnes Lab, Leiden
University, PObox 9504, 2300 RA Leiden, Netherlands}
\date{\today}
\begin{abstract}
A transition from local to global shear zones is reported for
granular flows in a modified Couette cell. The experimental geometry
is a slowly rotating drum which has a stationary disc of radius
$R_s$ fixed at its bottom. Granular material, which fills this cell
up to height $H$, forms a wide shear zone which emanates from the
discontinuity at the stationary discs edge. For shallow layers
$(H/R_s \alt 0.55)$, the shear zone reaches the free surface, with
the core of the material resting on the disc and remaining
stationary. In contrast, for deep layers $(H/R_s \agt 0.55)$, the
shear zones meet below the surface and the core starts to precess. A
change in the symmetry of the surface velocities reveals that this
behavior is associated with a transition from a local to a global
shear mode.
%
%
%
%
\end{abstract}

\vspace{-0.1mm} 

\maketitle

Slowly sheared granular matter usually organizes into rigid regions
separated by narrow shear bands where the material yields and flows
\cite{gdr,nedderman,oda,bridgewater,mulhaus,mueth,howell,losert1}. Such
grain flows appear to prohibit successful continuum modeling -- not
only because of the steep gradients in velocity, but also because
subtle microscopic characteristics of the granulate can alter the flow
in a qualitative manner \cite{mueth}.

\begin{figure}[t]
\includegraphics[width=8.cm]{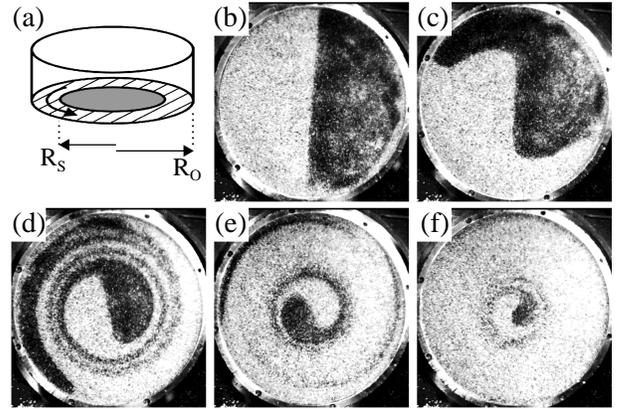}
\caption{(a) Schematic side-view of our split-bottomed shear cell,
showing a stationary bottom disk of radius $R_s$ (grey), the
rotating bottom ring (striped) and rotating outer cylinder of radius
$R_o\!=\!105$ mm. (b-f) Series of snapshots of top-views of our
setup (for $R_s\!=\!95$ mm and $H\!=\!60$ mm), where colored
particles sprinkled on the surface illustrate the core precession
for $t\!=\!0$ s (b), $t\!=\!10$ s (c), $t\!=\!10^2$ s (d),
$t\!=\!10^3$ s (e) and $t\!=\!10^4$ s (f).
 }\label{fig1a}
\end{figure}

The formation of narrow shear bands can be avoided by driving the
granulate from a discontinuity in the bottom support of the grain
layer (Fig.~\ref{fig1a}a), effectively pinning a wide shear zone
away from the sidewalls \cite{fenistein1}. The resulting
grain flows are smooth and robust, with both velocity profiles and
shear zone location exhibiting simple, grain independent properties
-- these flows should be amenable to a continuum description. At
present, a simple model aimed at describing an infinitely narrow
shear band captures the location of the shear zone \cite{unger}, but
there are no models which describe the finite width or velocity
profile of the shear zones found in experiment. In this Letter we
uncover a novel transition of the flow in this system which occurs
for deep grain layers.

The flows of {\em shallow} layers of granular materials in our setup
are characterized by wide shear zones. The surface velocities are
azimuthal and proportional to the driving rate $\Omega$, and the ratio
of the average angular surface velocity and $\Omega$, denoted
$\omega(r)$, is well fitted by \cite{fenistein1}:
\begin{equation}\label{2parameter}
\omega(r)= \mbox{nerf}~(\frac{r-R_c}{W})~,
\end{equation}
where $r$ is the radial coordinate, $R_c$ and $W$ parameterize the
location and width of the shear zones, and nerf$(x)$ denotes the
normalized errorfunction $1/2+1/2~ \mbox{errorf}(x)$.  $W$ is
independent of disc radius $R_s$ and grows with grain diameter $d$ and
layer height $H$ as \cite{fenistein1}
\begin{equation}\label{w_eq}
W/d \sim (H/d)^{2/3} \leftrightarrow W \sim H^{2/3}d^{1/3}~,
\end{equation}
while $R_c$ is independent of $d$ \cite{fenistein1}:
\begin{equation}\label{5/2}
1-R_c/R_s = (H/R_s)^{5/2}~.
\end{equation}
When $H/R_s$ is small, $\omega$ tends to zero for small $r$ -- the
core material remains {\em stationary} in shallow layers. The inward
shift of $R_c$  with $H/R_s$ (Eq.~(\ref{5/2})) implies, however,
that this behavior should break down eventually.

Figs.~\ref{fig1a}b-f illustrate novel flow patterns, characteristic
for {\em deep} layers. The most striking feature is that the core
starts to precess with a nonzero rate $\omega_p\!=\!
\omega(r\!\rightarrow\!0)$. Precession is not simply the consequence
of the overlap of the two opposing shear zones given by Eq.~(1) at
$r=\pm R_c$, since before being eroded by shear,
 the inner core rotates as a solid blob for an
appreciable time (Fig.~1(b-f)).

In this Letter we will characterize the transition to precession,
and specifically address the following questions: How does
$\omega_p$ grow with layer depth? What are the velocity profiles for
such deep layers? What happens to the scaling relations Eq.~2-3?
What underlies this transition?

%

\begin{figure}[t]
\includegraphics[width=8.0cm]{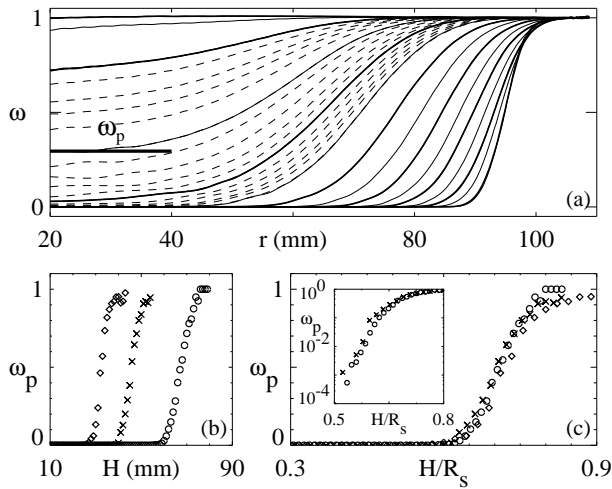}
\caption{(a) Surface velocity profiles $\omega(r)$ for $R_s\!=\!95$
mm and increasing layer depth. Thick curves: $H\!=\! 10,20,\dots,80$
mm; Thin curves $H\!=\! 15,25,\dots75$ mm; Dashed curves
$H\!=\!56,56,\dots 69$ mm. Precession gradually sets in for
 $H \agt 60$ mm. The thick line indicates the nonzero plateaux
 value $\omega_p$ for $H\!=\!65$ mm. (b)
  Precession rate $\omega_p$ as function of $H$ for
$R_s\!=\!45$ mm (diamonds), $R_s\!=\!65$ mm (x) and $R_s\!=\!95$ mm
(circles). (c) Precession rate curves collapse when plotted as
function of $H/R_s$. The inset shows that $\omega_p$ grows smoothly
with $H/R_s$.  }\label{fig2}
\end{figure}

{\em Setup --} Our setup is a modified version of the disk geometry
described in earlier work \cite{fenistein1}, and consists of a
stationary bottom disk of radius $R_s$, a rotating bottom ring and
outer cylinder of radius $R_o\!=\!105$ mm (Fig.~1a). The disc radius
$R_s$ can be varied from 45 mm to 95 mm. The cell is filled to a
height $H$ with a polydisperse mixture of spherical glass beads with
diameters ranging from 0.6 to 0.8 mm; a layer of grains is glued to
the side walls and bottom rings to obtain rough boundaries. Slow flows
are rate-independent, i.e., surface velocities are proportional to the
rotation rate $\Omega$ of outer cylinder and its co-moving ring. We
fix $\Omega$ at $0.15$ rad/s, and recover the surface velocity by a
variant of particle image velocimetry (for more details see
\cite{fenistein1}).

\begin{figure}[t]
\includegraphics[width=5.6cm]{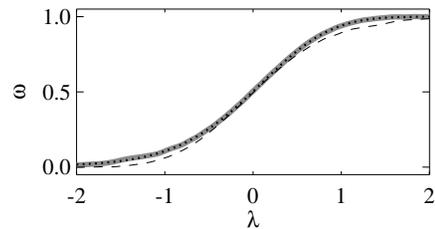}
\caption{Illustration of broken left-right symmetry
 of $\omega$ for $R_s=95$ mm and $H\!=\!58$ mm, where $\lambda$ denotes the rescaled spatial
 coordinate $((r-R_c)/R_s)$.
 Shown are $\omega(\lambda)$ (grey), its
  symmetric counterpart $1-\omega(-\lambda)$
 (dashed) and the best fit of
 $\omega(\lambda)$ to Eq.~(\ref{cubic}) (dots).}\label{fig3}
\end{figure}

{\em Basic phenomenology -- } Figs.~\ref{fig2}-\ref{fig3} illustrate
the main features of the surface velocity profiles $\omega(r)$ as
function of increasing layer depth $H$. For shallow layers
($H/R_s\alt0.5$) the data is well fitted by Eq.~\ref{2parameter}.
The core precession, which sets in for $H/R_s \alt 0.65$ constitutes
the clearest deviation from this simple form. For $H/R_s\!
\agt\!0.8$ $\omega_p$ tends to one, and the whole surface rotates
then rigidly. Apparently the shear zone remains confined below the
surface in this case.

In Fig.~\ref{fig2}b-c we show that the data for $\omega_p$ collapses
when plotted as function of $H/R_s$. The data  suggests a much
faster than exponential growth in the initial stages (i.e. when
$\omega_p \ll 1$) as illustrated in the inset of Fig.~\ref{fig2}c).
Before precession becomes appreciable, the left-right symmetry of
$\omega(r)$ is broken: $\omega(\Delta r -R_c) \neq
1-\omega(R_c-\Delta r)$. As illustrated in Fig.~\ref{fig3}, the
right (large $r$) tail of $\omega(r)$ is then significantly steeper
than its left (small $r$) tail.

Hence, three regimes can be distinguished. {\em Shallow layers}
occur for $H/R_s\!\alt\!0.45$ where $\omega_p$ is zero, $\omega(r)$
is symmetric and Eq.~\ref{2parameter} fits the data well. {\em Deep
layers} occur for $H\!\agt\!0.65$, where the reflection symmetry of
zones is strongly broken and precession sets in. In the {\em
crossover regime}, $0.45\!\alt\!H\!\alt\!0.65$, the
 symmetry of $\omega(r)$ is broken but $\omega_p$ remains small.

\begin{figure}[t]
\includegraphics[width=8.cm]{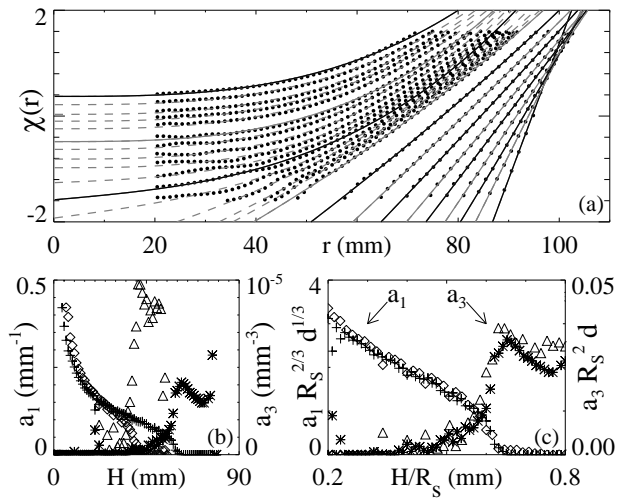}
\caption{ (a) Profiles of $\chi(r)$ for $R_s\!=\!95$ mm and
increasing layer depth (dots), compared to cubic fits given by
Eq.~(\ref{cubic}) (curves). Similar to  Fig.~2a, $H\!=\!
10,15,20,\dots,55,56,\dots,70$ mm. (b) Fit parameters $a_1$ for
$R_s\!=\!65$ mm (diamonds) and 95 mm (+) and $a_3$ for $R_s\!=\!65$
mm (triangles) and 95 mm (stars) as function of $H$. (c)
Non-dimensionalized fit parameters as function of $H/R_s$ (see
text).
 }\label{fig4}
\end{figure}

{\em Transition from local to global modes --} Both symmetry breaking
and precession reflect the same underlying change in the qualitative
nature of the shear modes. Since $\omega(r)$ in general can be written
as ${\mbox{nerf}}\left[ \chi(r)\right]$, we have calculated $\chi(r)
:= {\mbox{nerf}}^{-1} \left[\omega(r) \right]$ from our experimentally
obtained velocity profiles. The resulting $\chi(r)$ are shown in
Fig.~\ref{fig4}a. For shallow layers $\chi(r)$ is a linear function,
consistent with the fact that $\omega(r)$ obeys
Eq.~(\ref{2parameter}). For deep layers, $\chi(r)$ becomes a nonlinear
function and over the whole range of $H$, $\chi(r)$ can be fitted well
by an empirical third order polynomial \cite{quadnote} (see
Fig.~\ref{fig4}a):
\begin{equation}\label{cubic}
\chi(r) = a_0 + a_1 r + a_3 r^3~.
\end{equation}
{\em Both precession and symmetry breaking are well captured by this
fit.}  As shown in Fig.~\ref{fig4}b-c, the three aforementioned
regimes have now a simple interpretation in terms of the linear
$(a_1)$ and cubic $(a_3)$ coefficient. For shallow layers the cubic
term vanishes while for deep layers the linear term is absent; both
are nonzero in the crossover regime.

Note that the fit parameters $a_1$ and $a_3$ have dimensions of
inverse length and inverse length cubed, and are therefore expected
to scale as $W$ and $W^3$. Rewriting Eq.~\ref{w_eq} as
$W/(R_s^{2/3}) d^{1/3} \sim (H/R_s)^{2/3}$, it follows that $a_1
R_s^{2/3} d^{1/3}$ and $a_3 R_s^2 d$ should collapse when plotted as
function of $H/R_s$ -- indeed this is the case \cite{dnote} (Fig.~4c).

The functional form for $\omega$ for deep layers is thus
\begin{equation}\label{deep}
\omega(r) \approx \mbox{nerf}(a_0 + a_3 r^3)~,
\end{equation}
while for shallow layers Eq.~(\ref{2parameter}) can be rewritten as
\begin{equation}\label{shallow}
\omega(r) \approx \mbox{nerf}(a_0 + a_1 r)~.
\end{equation}
These two limits have different symmetries. For shallow layers, two
flow profiles obtained for two different slip radii ($R_{s1}$ and
$R_{s2}$) are simply related by translations of the radial
coordinate via
$\omega(r-R_{s1})|_{R_{s1},H}=\omega(r-R_{s2})|_{R_{s2},H}$ -- the
shear zone is a {\em local} phenomenon, insensitive to the location
of the center of the shear cell. For deep layers this symmetry is
absent, and the shear mode is {\em global}. The precession and
symmetry breaking of $\omega(r)$ both reflect the crossover from a
local to a global shear mode.

{\em 3D flow structure --} In Fig.~5, the crucial difference between
the 3D flow structure of shallow and deep layers is illustrated. The
flow in the bulk has been probed by putting patterns of lines of
colored tracer particles at given height in the bulk, $H_b$, adding
more material so as to bury the line-pattern, rotating the system
for a short period ($\sim 8$ s), and recovering the deformed
line-pattern by carefully removing the upper layers of grains
\cite{fenistein1}. The striking difference between these two cases
where $\omega_p\! \approx \!0$ and a  $\omega_p\! \approx \!0.8$
respectively, is that in the former case the shear zones reach the
free surface while for the latter case the shear zones meet in the
bulk. This scenario is consistent with the finding of recent MRI and
numerical studies of the flow in this geometry \cite{paul,chang},
and gives a straightforward interpretation of local and a global
shear modes.

\begin{figure}[t]
\includegraphics[width=7.5cm]{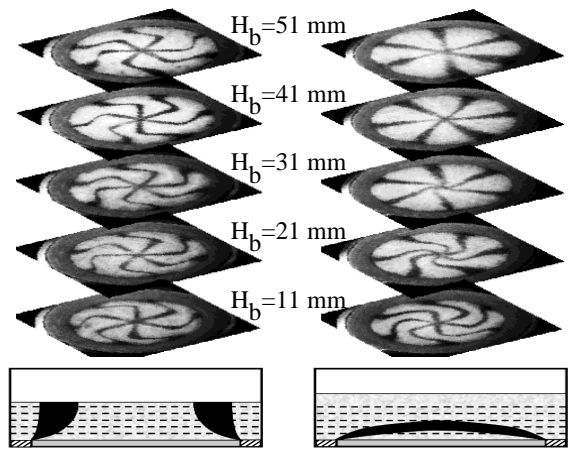}
\caption{3D flow-profiles for $R_s\!=\!95$ mm. Both stacks show five
slices at heights $H_b=11,21,\dots,51$ mm inside the material; the
left stack is for layer depth $H=57$ mm, the right stack for $H=71$
mm. The bottom figures show sketches of the shear zones within the
bulk (black) -- the dashed curves indicate the heights $H_b$ where
the patterns where created.
 }\label{fig6}
\end{figure}

\begin{figure}[b]
\includegraphics[width=8.cm]{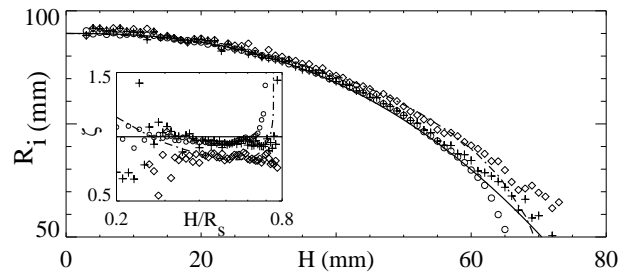}
\caption{Shear zone positions versus layer depth for $R_s\!=\!95$
mm, where circles, pluses and diamonds correspond to $R_1, R_2$ and
$R_3$ respectively, the solid curve is the scaling form given by
Eq.~(\ref{5/2}) and the dashed curve is the result by Unger {\em et
al.} \cite{unger}. The inset shows the same data, now in rescaled
form, by defining $\zeta:=(1-R_i/R_s)/(H/R_s)^{5/2}$. }\label{fig5b}
\end{figure}

{\em Comparison to theory -- } Recently, Unger {\em et al.} proposed
a simple theory for the location of the shear zone in our system.
They model the shear zone as an infinitely thin sheet along which
the granulate slides, with the normal stress on the sheet given by
hydrostatic pressure and the shear stress by sliding friction. By
numerically adjusting the sheet so as to minimize the total torque,
the theory predicts the shear zones location at the free surface,
$R_m(H)$. This function is different from the simple scaling law
Eq.~(\ref{5/2}), and the theory predicts that for deep layers the
shear sheet no longer reaches the free surface but closes in the
bulk; the two cases are separated by a first order (hysteretic)
transition that occurs around $H/R_s \approx 0.7$ \cite{unger}.
While we have not found any evidence for hysteresis, the theory of
Unger {\em et al.} does a good job in capturing the qualitative
behavior.

To test the theory quantitatively, in Fig.~6 our data for the shear
zones location is compared to $R_m(H)$ and Eq.~(\ref{5/2}). For the
general case, we have to define the center of general shear zones.
There is no unique choice, and we have tested the following three
expressions:
\begin{eqnarray}
R_1 &:=& \mbox {where $\omega(r) =0.5$}~,\\
R_2 &:=& \mbox {where $\partial_r \omega(r)$ is maximal}~,\\
R_3 &:=& \mbox {where $\partial_r (r \omega(r))$ is maximal}~.
\end{eqnarray}
Note that for shear zones of the form Eq.~({\ref{2parameter}), $R_1$
and $R_2$ coincide with $R_c$, while $R_3$ corresponds to where the
dimensional strain rate is maximal. As shown in Fig.~\ref{fig5b},
for shallow layers, both $R_1$ and $R_2$ are better described by
Eq.~(\ref{5/2}) than by $R_m$, while $R_3$ appears to be better
described by $R_m$. For deep layers, the situation is more
complicated, with neither model describing any of the three measured
curves in detail.

{\em Open Questions -- } There are three questions that
we think deserve particular attention. The first concerns the
functional form of $\omega(r)$. While it is now widely accepted that
in many shear flows, the velocity profile across a shear band is
smooth, it is not clear what mechanism selects this velocity profile,
or indeed its ``tail'' \cite{mueth}. In some systems, these tails
appear exponential \cite{komatsu}, for our shallow shear zones these
tails are Gaussian \cite{fenistein1}, and our results here suggest
that for deep layers, the tail stretching out into the material has a
more complex form (Eq.~(\ref{deep})). Similarly, we do not know what
determines the functional form or tails of $\omega_p(H/R_s)$
(Fig.~2b-c).  Theoretical work is very welcome here.

\begin{figure}[t]
\includegraphics[width=8.cm]{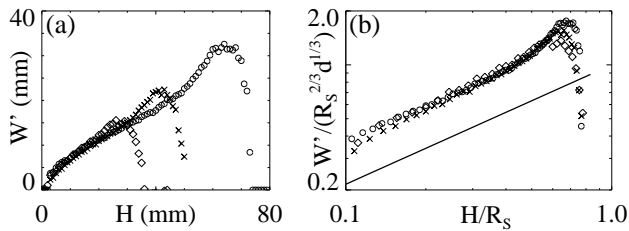}
\caption{ (a) Shear zones width $W'$ as function of $H$ for $R_s\!=$
45, 65 and 95 mm. For small heights the shearzone width is
independent of $R_s$, while deviations start to set in for larger
heights.  (c) The nondimensionalized width $W' R_s^{2/3} d^{1/3}$ follows
a universal curve as function of $H/R_s$, which for small heights is
approximated well by a powerlaw with exponent 2/3, as indicated by the
straight line.}\label{fig5a}
\end{figure}

The second question concerns the finite width of the shear zones. Let
us define the width $W'$, for general velocity profiles, as the
interval where $\omega$ grows from $\omega_p+0.1\times(1-\omega_p)$ to
$0.9\times(1-\omega_p)$ \cite{wnote}.  $W'$ does not depend on $R_s$
for shallow layers, but $R_s$ sets the scale where $W'$ saturates; for
deep layers, both the microscale ($d$) and macroscale ($R_s$) play a
role (Fig.~7a). The ratio $R_s/d$ between these scales can be
conceived as the non-dimensionalized radius of curvature.
Eq.~(\ref{w_eq}), when rewritten as $W'/d=(H/R_s)^{2/3}(R_s/d)^{2/3}$,
suggests to plot the rescaled width, $W'/(R_s^{2/3} d^{1/3})$, as
function of $H/R_s$ -- as shown in (Fig.~7b), this leads to a good
data-collapse. This result is consistent with the scaling with
exponent 2/3 for shallow layers (Eq.~(\ref{w_eq})). It is not known
what sets the value of this non-trivial exponent.

Finally, it is an open question whether the transition to precession
should be conceived as a smooth crossover or as a sharply defined
transition. The smooth growth of $\omega_p$ with $H/R_s$ suggests a
crossover (Fig.~2b-c), while the (critical) vanishing of $a_1$ with
$(H/R_s)$ (Fig.~4b-c) suggests a sharp transition.

{\em Outlook -- } A wide range of granular flows can be achieved in
split bottom geometries. Detailed studies of the 3D structure of these
flows \cite{chang}, together with studies of the role of packing
density \cite{paul}, anisotropies \cite{Behringer} and contact forces
\cite{corwin} are uncovering the richness of slowly sheared
granulates, and provide crucial ingredients for theories of flows of
dense granular media. The robust behavior of grain flows in split
bottom geometries hopefully will provide important testing ground for
such theories \cite{unger,depken}.


{\em Acknowledgments} We gratefully acknowledge discussions with X.
Chang, H. Jaeger, S. Nagel, T. Unger, J. Kertesz, M. Depken and W.
van Saarloos, and financial support from ``Nederlandse Organisatie
voor Wetenschappelijk Onderzoek (NWO)'' and  ``Stichting
Fundamenteel Onderzoek der Materie (FOM)''.


\begin{thebibliography}{99}

\bibitem{gdr} GDR MiDi, Eur. Phys. J. E {\bf 14}, 341 (2004)

\bibitem{nedderman} R. Nedderman, \textit{Statics and Kinematics of
Granular Materials} (Cambridge University Press 1992).

\bibitem{oda} M. Oda and H. Kazama, G\'{e}othechnique \textbf{48}, 465 (1998).

\bibitem{bridgewater} J. Bridgewater, G\'{e}othechnique \textbf{30}, 533 (1980).
\bibitem{mulhaus} H. B. Muhlhaus and I. Vardoulakis, G\'{e}otechnique
\textbf{37}, 271 (1987).

\bibitem{mueth} D.M. Mueth, \textit{et al.} Nature \textbf{406}, 385
  (2000); D.M. Mueth, Phys. Rev. E {\bf 67}, 011304 (2003).

\bibitem{howell} D. W. Howell, R.P. Behringer and C.T. Veje,
Phys. Rev. Lett. \textbf{82}, 5241 (1999).

\bibitem{losert1} W. Losert, L. Bocquet, T.C. Lubensky and
J.P. Gollub, Phys. Rev. Lett. \textbf{85}, 1428 (2000); W. Losert and
G. Kwon, Advances in Complex systems \textbf{4}, 369 (2001); M. Toiya,
J. Stambaugh, and W. Losert, Phys. Rev. Lett. {\bf93}, 088001 (2004)

\bibitem{fenistein1} D. Fenistein and M. van Hecke, Nature {\bf 425},
256 (2003); D. Fenistein, J.-W. van de Meent and M. van Hecke,
Phys. Rev. Lett. {\bf92}, 094301 (2004).

\bibitem{unger} T. Unger, J. T\"or\"ok, J. Kert\'esz and D. E. Wolf,
Phys. Rev. Lett. {\bf 92}, 214301 (2004).

\bibitem{quadnote} The more obvious choice $\chi(r) \approx a_0 + a_1
r + a_2 r^2$ fails to fit the data for deep layers.

\bibitem{dnote} As microscopic scale we take $d\!=\!0.7$ mm.

\bibitem{paul} P. Umbanhowar, private communications.

\bibitem{chang} X. Chang {\em et al.}, submitted (2005).

\bibitem{komatsu} T. S. Komatsu, S. Inagaki, N. Nakagawa and
S. Nasuno, Phys. Rev. Lett. \textbf{86}, 1757 (2001).

\bibitem{wnote} For shallow layers, $W'\!\approx1.812W$ (see
Eq.~\ref{2parameter}).

\bibitem{Behringer} T. S.  Majmudar and R. P. Behringer,
Nature {\bf 435}, 1079 (2005)

\bibitem{corwin} E.I. Corwin, H.M. Jaeger and S.R. Nagel, Nature {\bf
435}, 1075 (2005).

\bibitem{depken} M. Depken, M. van Hecke and W. van Saarloos, in
preparation.

\end{thebibliography}
\end{document}